\documentclass[twocolumn,amsmath,amssymb,aps,pra]{revtex4-2}

\usepackage{graphicx}
\usepackage{dcolumn}
\usepackage{bm}
\usepackage{color}

\begin{document}

\title{Measurement-free fault-tolerant logical zero-state encoding of the distance-three nine-qubit surface code in a one-dimensional qubit array}

\author{Hayato Goto,  Yinghao Ho, and Taro Kanao}
\affiliation{
Frontier Research Laboratory, 
Corporate Research \& Development Center, 
Toshiba Corporation, 
1, Komukai Toshiba-cho, Saiwai-ku, Kawasaki-shi, 212-8582, Japan}

\date{\today}

\begin{abstract}

Generation of logical zero states encoded with a quantum error-correcting code is the first step for 
fault-tolerant quantum computation, but requires considerably large resource overheads in general. 
To reduce such overheads, we propose an efficient encoding method for the distance-three, nine-qubit surface code and show its fault tolerance. 
This method needs no measurement, unlike other fault-tolerant encoding methods.
Moreover, this is applicable to a one-dimensional qubit array. 
Observing these facts, we experimentally demonstrate the logical zero-state encoding of the surface code using a superconducting quantum computer on the cloud. 
We also experimentally demonstrate the suppression of fast dephasing due to intrinsic residual interactions in this machine 
by a dynamical decoupling technique dedicated for the qubit array. 
To extend this method to larger codes, we also investigate the concatenation of the surface code with itself, 
resulting in a distance-nine, 81-qubit code.
We numerically show that fault-tolerant encoding of this large code can be achieved by appropriate error detection. 
Thus, the proposed encoding method will provide a new way to low-overhead fault-tolerant quantum computation.

\end{abstract}

\maketitle

\section{Introduction}

Quantum computers are notoriously prone to errors due to decoherence and nonideal gate operations.
To suppress these errors and perform quantum algorithms for, e.g., prime factoring~\cite{Shor1994a,Ekert1996a} and 
quantum chemistry calculations~\cite{Aspuru2005a,McArdle2020a}, 
fault-tolerant quantum computation (FTQC) using quantum error-correcting codes~\cite{Nielsen,Shor1996a,Gottesman1998a,Knill1998a} is highly expected. 
However, the FTQC in general requires large computational resource overheads~\cite{Jones2012a,Fowler2012a,Gidney2021a}, 
which is the case even for the first step in FTQC, namely, 
preparation of logical zero states encoded with a quantum error-correcting code.
Hence, the reduction of such overheads for logical-qubit state preparation is highly desirable.

For example, an efficient fault-tolerant method for preparation of logical qubit states has been proposed 
for the Steane seven-qubit code (one of the smallest distance-three codes capable of correcting arbitrary single-qubit errors)~\cite{Goto2016a}, 
which has enabled experimental realization of the Steane-code logical qubit states, 
including so-called magic states~\cite{Bravyi2005a,Reichardt2005a,Goto2014a,Chamberland2019a}, 
using laser-cooled trapped ions~\cite{Ryan2021a,Postler2022}.
However, the method still requires an ancilla qubit and its measurement to achieve its fault tolerance.

In this paper, we propose an efficient fault-tolerant logical zero-state encoding method for the nine-qubit surface code~\cite{Horsman2012a,Tomita2014a}, 
which is another small distance-three code.
Remarkably, this method needs no ancilla qubit and no measurement~\cite{comment,Satzinger2021a}, 
which is in contrast to the conventional surface-code encoding method repeating syndrome measurements 
realized in recent experiments~\cite{Zhao2022a,Krinner2022a,Google2023a}. 
Moreover, the proposed method is applicable to a one-dimensional qubit array, which allows us to experimentally demonstrate this method 
using a superconducting quantum computer on the cloud.
(Interestingly, a measurement-free fault-tolerant zero-state encoding method has recently been proposed 
for the Steane code~\cite{Heuben2023a}. 
Unlike the proposed method, however, this needs two ancilla qubits and a Toffoli gate, and also cannot be realized in a one-dimensional qubit array.)

To extend the proposed method to larger codes, 
we also investigate the distance-nine, 81-qubit code obtained by concatenating the nine-qubit surface code with itself.
We numerically show that fault-tolerant logical zero-state encoding of this large code can be achieved by 
three level-1 error-detecting teleportations~\cite{Knill2005a,Goto2009a,Goto2013a}, 
where soft-decision decoding based on 
conditional probability calculations~\cite{Goto2013a,Poulin2006a,Goto2014b} is essential for high performance.
Because of the assumption of arbitrary two-qubit gates, 
the experimental realization of this encoding is challenging 
for superconducting quantum computers, 
but will be possible for recently developed neutral-atom quantum computers using optical tweezers~\cite{Bluvstein2022a,Graham2022a}.

\section{Measurement-free fault-tolerant logical zero-state encoding of the nine-qubit surface code}
\label{sec-encoder}

\begin{widetext}

\begin{figure}[t]
	\includegraphics[width=16cm]{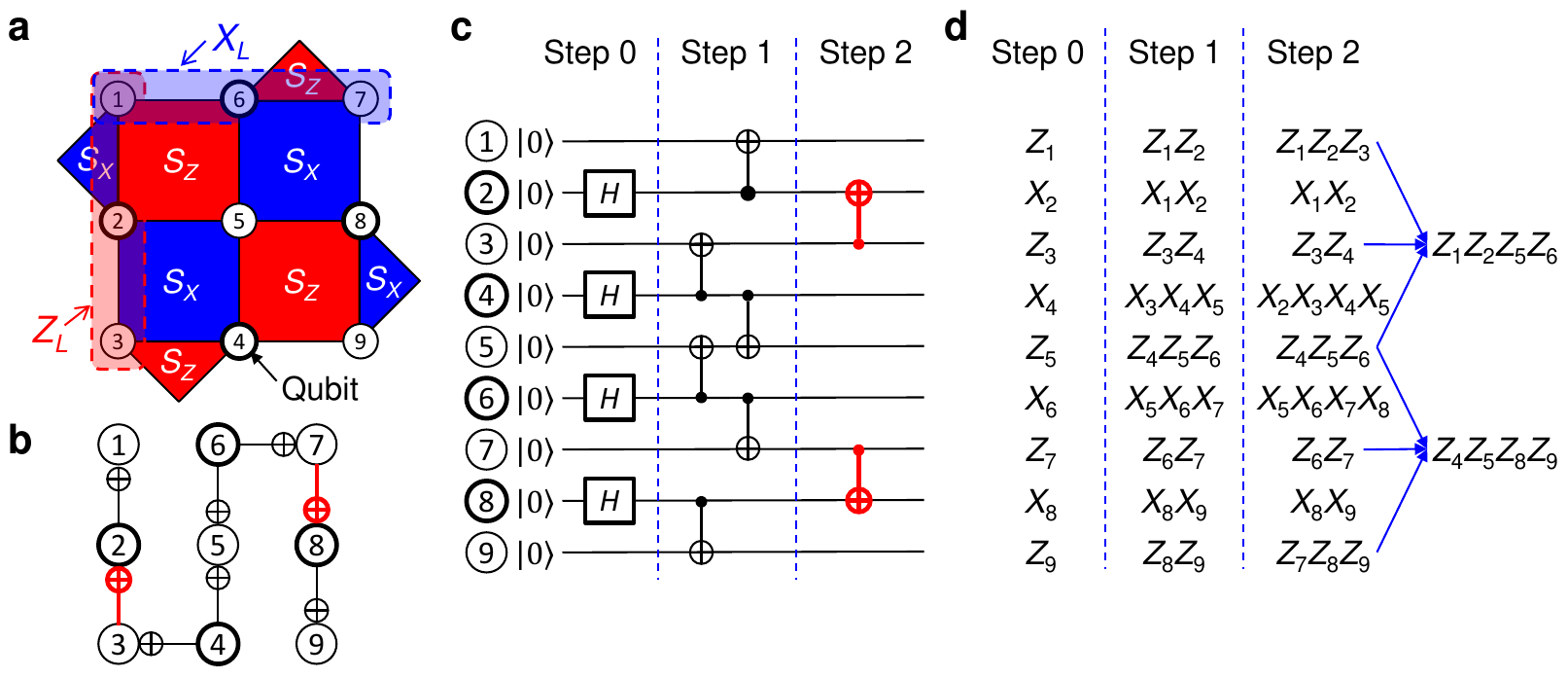}
	\caption{\textbf{Measurement-free fault-tolerant logical zero-state encoding of 
	the nine-qubit surface code.}
	\textbf{a}~Definition of the nine-qubit surface code.
	$S_Z$ and $S_X$ denote its $Z$ stabilizers 
	($Z_6 Z_7$, $Z_1 Z_2 Z_5 Z_6$, $Z_4 Z_5 Z_8 Z_9$, $Z_3 Z_4$) and $X$ stabilizers 
	($X_1 X_2$, $X_2 X_3 X_4 X_5$, $X_5 X_6 X_7 X_8$, $X_8 X_9$), respectively.
	${Z_L=Z_1 Z_2 Z_3}$ and ${X_L=X_1 X_6 X_7}$ are its logical Pauli operators.
	The logical zero state $|0\rangle_L$ is the simultaneous eigenstate of the eight stabilizers and $Z_L$ 
	with eigenvalues of 1.
	\textbf{b}~Graphical description of the proposed method.
	Bold circles indicate that the corresponding qubits are initialized to 
	${|+\rangle = H|0\rangle = (|0\rangle + |1\rangle)/\sqrt{2}}$, 
	where $H$ denotes the Hadamard gate, and the others are to $|0\rangle$.
	Two bold CNOT gates in red are performed after the other six CNOT gates.
	\textbf{c}~Quantum circuit corresponding to \textbf{b}.
	\textbf{d}~Nine stabilizers describing the nine-qubit state at the end of each step in \textbf{c}.}
	\label{fig-encoder}
\end{figure}

\end{widetext}

The proposed encoding method for the nine-qubit surface code is summarized in Fig.~\ref{fig-encoder}, 
where Fig.~\ref{fig-encoder}\textbf{a} shows the definition of the code, 
Figs.~\ref{fig-encoder}\textbf{b} and \ref{fig-encoder}\textbf{c} show the proposed method, 
and Fig.~\ref{fig-encoder}\textbf{d} shows how the stabilizers of the nine-qubit state change during the encoding~\cite{comment-stabilizer}.
The final stabilizers in Fig.~\ref{fig-encoder}\textbf{d} are exactly the same as those of the logical zero state, $|0\rangle_L$, 
encoded with the nine-qubit surface code (see Fig.~\ref{fig-encoder}\textbf{a}). 
This means that the proposed method can successfully encode $|0\rangle_L$.
Also note that this encoding needs no ancilla qubit and no measurement, as mentioned above.

The fault tolerance of the method is explained as follows.
At the end of Step~1 in Fig.~\ref{fig-encoder}\textbf{c}, 
the weights (the numbers of non-identity Pauli operators) of all the nine stabilizers are two or three, 
as shown in Fig.~\ref{fig-encoder}\textbf{d}, 
which allows us to regard any correlated $X$ and $Z$ errors induced by 
the six controlled-NOT (CNOT) gates in Step~1 
as single-qubit $X$ and $Z$ errors, respectively.
(In this work, we assume that correlated errors occur only by two-qubit gates.)
Hence, we have only to care correlated errors induced by the two CNOT gates in Step~2.
(A similar technique is used for the above-mentioned method for the Steane code~\cite{Goto2016a}.)
The correlated errors are $X_2 X_3$ and $X_7 X_8$, 
which can be corrected by this code~\cite{comment-correction}. 
Thus, this encoding induces no uncorrelatable errors up to the first order of physical error rates, 
which means its fault tolerance.

\section{Experimental demonstration of the proposed surface-code encoding}
\label{sec-experiment}

\begin{figure}[b]
	\includegraphics[width=8.5cm]{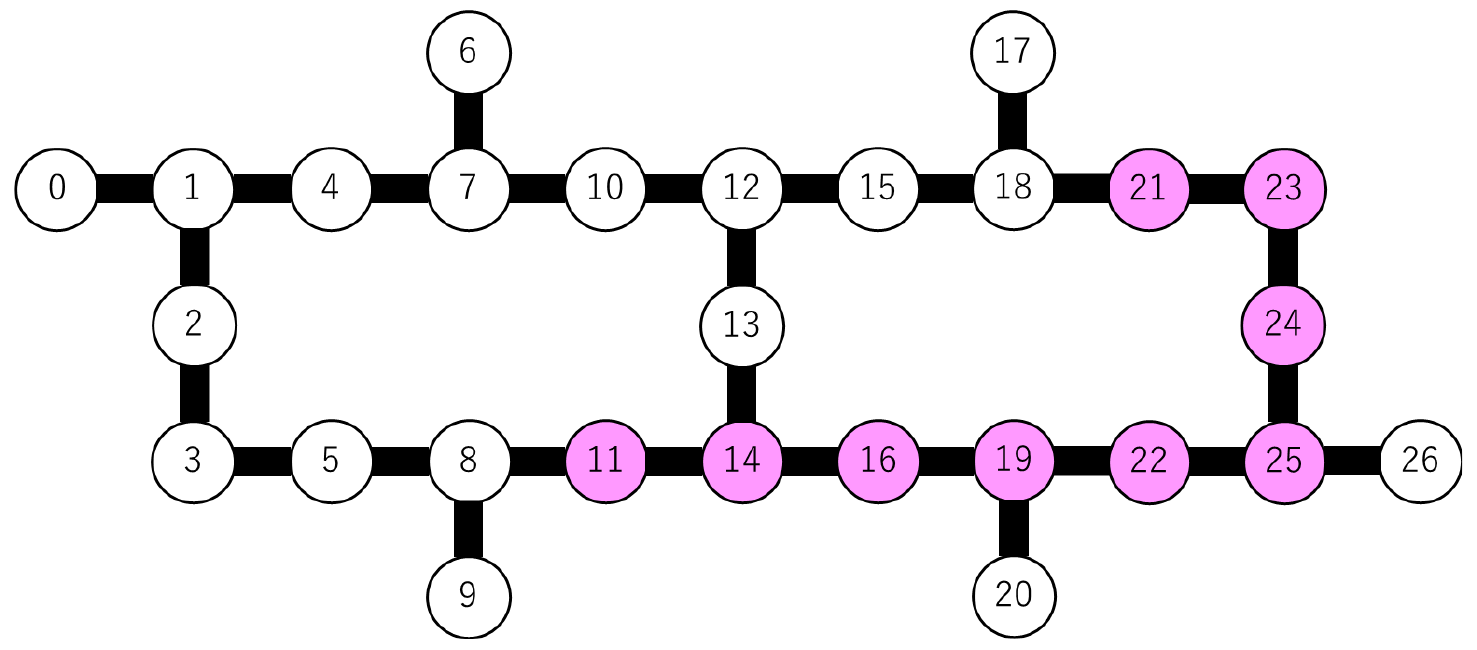}
	\caption{\textbf{Qubit layout of the superconducting quantum computer used in this work.}
	This is 27-qubit ibm-kawasaki~\cite{IBM}.
	Qubits 21, 23, 24, 25, 22, 19, 16, 14, and 11 in this machine are used for Qubits 1--9 in Fig.~\ref{fig-encoder}.}
	\label{fig-IBM}
\end{figure}

Observing that the proposed encoding can be realized using a one-dimensional qubit array, 
as shown in Fig.~\ref{fig-encoder}\textbf{b}, 
we did experiments of this encoding using a superconducting quantum computer on the cloud 
(27-qubit ibm-kawasaki~\cite{IBM}), the qubit layout of which is shown in Fig.~\ref{fig-IBM}.
We used nine qubits in a row among the 27 qubits, which are highlighted in Fig.~\ref{fig-IBM}.
The performance of this machine is summarized in Tables~\ref{table-CNOT} and \ref{table-coherence}.

First, we prepare $|0\rangle_L$ according to the quantum circuit in Fig.~\ref{fig-encoder}\textbf{c} 
and measure $Z$ of all the qubits after a delay time from the preparation.
We decode the measurement results using the hard-decision decoding rule in Appendix~\ref{appendix-hard-decoding}, 
and estimate the measurement result of $Z_L$.
If the decoding results in ${Z_L =-1}$, the decoding fails, which means a logical error.

\begin{table}[t]
	\caption{\textbf{CNOT-gate perfomance of 27-qubit ibm-kawasaki.}
	``CNOT" column shows the control-qubit (C) and target-qubit (T) numbers of this machine (see Fig.~\ref{fig-IBM}).}
	\begin{tabular}{ccc} \hline \hline
	CNOT~ & Error rate (\%) & Gate time (ns)
	\\ \hline
	C21, T23 & 0.51 & 292
	\\
	C23, T24 & 0.82 & 427
	\\
	C24, T25 & 0.58 & 363
	\\
	C25, T22 & 0.68 & 284
	\\
	C22, T19 & 0.88 & 281
	\\
	C19, T16 & 0.88 & 295
	\\
	C16, T14 & 0.54 & 295
	\\
	C14, T11 & 2.26 & 409
	\\ \hline \hline
	\end{tabular}
	\label{table-CNOT}
\end{table}

\begin{table}[t]
	\caption{\textbf{Qubit coherence times of 27-qubit ibm-kawasaki~\cite{comment-T2}.}
	``Qubit" column shows the qubit number of this machine (see Fig.~\ref{fig-IBM}).}
	\begin{tabular}{ccc} \hline \hline
	Qubit~ & $T_1$ ($\mu$s) & $T_2$ ($\mu$s)
	\\ \hline 
	Q21 & 144 & 108
	\\
	Q23 & 134 & 332
	\\
	Q24 & 169 & 125
	\\
	Q25 & 174 & 213
	\\
	Q22 & 147 & 188
	\\
	Q19 & 196 & 263
	\\
	Q16 & 128 & 133
	\\
	Q14 & 122 & 78
	\\
	Q11 & 113 & 110
	\\ \hline \hline
	\end{tabular}
	\label{table-coherence}
\end{table}

\begin{widetext}

\begin{figure}
	\includegraphics[width=16cm]{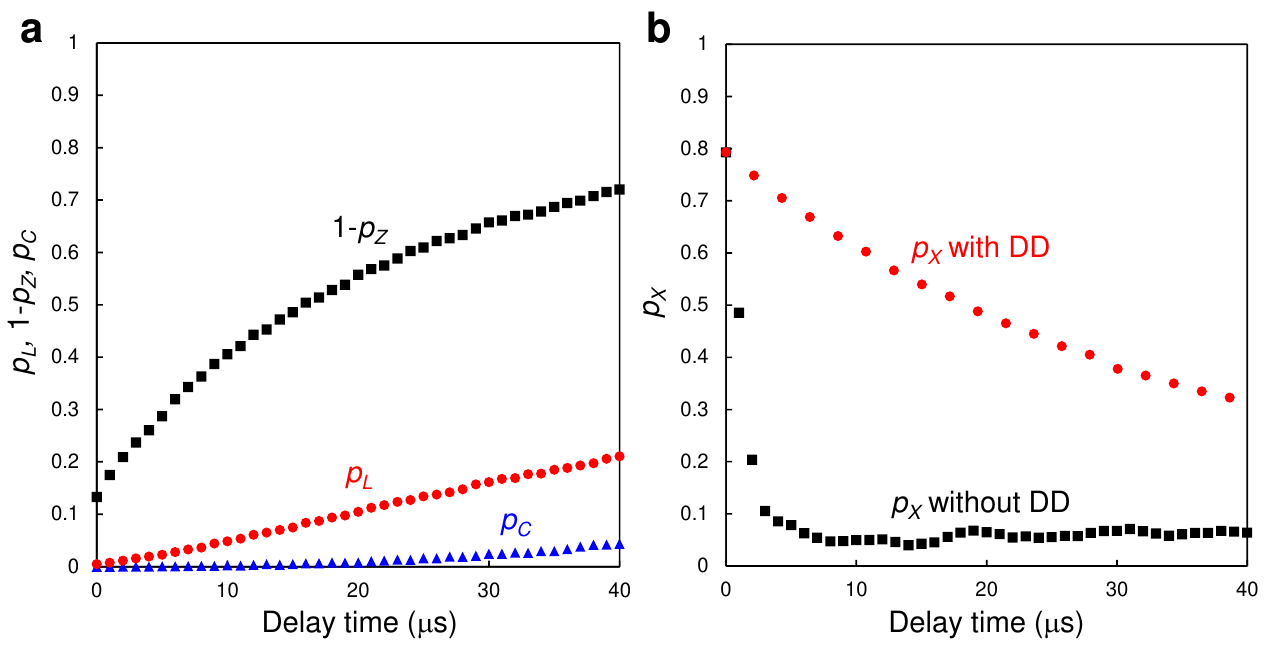}
	\caption{\textbf{Experimental results of the logical zero-state encoding of the nine-qubit surface code 
	using a superconducting quantum computer.}
	\textbf{a}~Delay-time dependence of the logical error probability $p_L$ (circles), 
	the conditional logical error probability $p_C$ (triangles) under the condition that all the four $Z$ stabilizers are 1,
	and the no-X-error probability $p_Z$ (squares) that both $Z_L$ and the four $Z$ stabilizers are 1. 
	\textbf{b}~Delay-time dependence of the no-$Z$-error probability $p_X$ 
	that all the four $X$ stabilizers are 1.
	The circles and squares show the results, respectively, with and without 
	the dynamical decoupling (DD) technique shown in Fig.~\ref{fig-DD}.
	In both \textbf{a} and \textbf{b}, statistical errors are negligible compared to symbol sizes.}
	\label{fig-experiment}
\end{figure}

\end{widetext}

The experimental results are shown in Figure~\ref{fig-experiment}\textbf{a}, 
where $p_L$ (circles) is the logical error probability, 
$p_C$ (triangles) is the conditional logical error probability under the condition that all the four $Z$ stabilizers are 1, and
$p_Z$ (squares) is the no-$X$-error probability that 
both $Z_L$ and the four $Z$ stabilizers are 1.
$p_L$ and $p_C$ are substantially lower than ${1-p_Z}$.
In particular, $p_L$ and $p_C$ are about 0.5\% and 0.004\%, respectively, at the zero delay time, 
which are, respectively, lower and much lower than physical-CNOT error rates (see Table~\ref{table-CNOT}).
These results indicates that $|0\rangle_L$ was successfully prepared and 
its logical error probability could be suppressed by error correction or detection.

Note that the above results depend only on $Z$ measurements (meaturements of bit-string states) and provide 
no information of $X$ measurements (quantum superpositions of bit-string states).
The logical zero state, $|0\rangle_L$, encoded with the nine-qubit surface code 
is a quantum superposition of 16 bit-string states satisfying the five $Z$-stabilizer conditions.
To evaluate the quantum superposition in the prepared $|0\rangle_L$, 
we also measured the $X$ stabilizers, the results of which are shown by squares in Fig.~\ref{fig-experiment}\textbf{b}.
When the delay time is zero, the no-$Z$-error probability, $p_X$, 
that all the four $X$ stabilizers are 1 is 79\%. 
(The corresponding value of $p_Z$ is 87\%, as found in Fig~\ref{fig-experiment}\textbf{a}.)
This result suggests that the prepared $|0\rangle_L$ was actually a quantum superposition of bit-string states, as expected. 
In fact, it is shown that the fidelity of the prepared $|0\rangle_L$ is lower bounded by ${p_Z + p_X - 1 = 0.66}$ 
(see Appendix~\ref{appendix-fidelity}), 
which is substantially higher than that for the maximally mixed state of the 16 bit-string states, namely, ${1/16=0.0625}$.
(This fidelity evaluation of $|0\rangle_L$ prepared by the proposed method would be useful 
for benchmarking of real quantum computers with at least nine qubits in a row.)

\section{Dynamical decoupling to suppress dephasing due to residual $ZZ$ interactions}
\label{sec-DD}

\begin{figure}[b]
	\includegraphics[width=7cm]{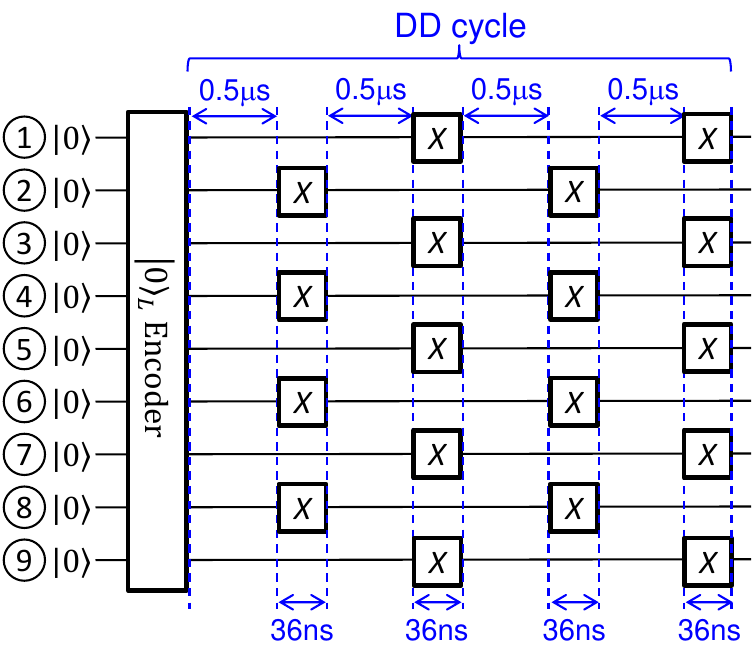}
	\caption{\textbf{Dynamical decoupling technique for suppressing dephasing in $|0\rangle_L$ due to $ZZ$ interactions.}
	Each dynamical decoupling (DD) cycle consists of four 0.5-$\mu$s delay times (more precisely, 0.5013~$\mu$s) followed 
	by four or five $X$ gates with gate times of 36~ns (more precisely, 35.6~ns).
	Thus, each DD cycle takes 2.1476~$\mu$s.
	It can easily be shown that this DD cycle can eliminate unwanted additional phase rotations due to $ZZ$ interactions.}
	\label{fig-DD}
\end{figure}

It is notable that $p_X$ decays much faster than $p_Z$. 
(The recovery of $p_X$ to ${1/2^4=0.0625}$ found in Fig.~\ref{fig-experiment}\textbf{b} 
can be explained by convergence to a maximally mixed state of the 16 bit-string states 
due to decoherence.)
Since this decay of $p_X$ is too fast compared to qubit coherence times (see Table~\ref{table-coherence}), 
the decay may be caused by another decoherence source.
We identify it with so-called residual $ZZ$ interactions inducing unwanted additional phase rotation 
only for $|11\rangle$ of adjacent qubits. 
To suppress this dephasing due to the $ZZ$ interactions, 
we applied a dynamical decoupling technique dedicated for the qubit array, 
which is shown in Fig.~\ref{fig-DD}.
(Similar studies with different techniques have been reported~\cite{Pokharel2018a,Tripathi2022a}.)
The results are shown by circles in Fig.~\ref{fig-experiment}\textbf{b}, 
which clearly shows that the decay of $p_X$ becomes substantially slower, as expected.
This concludes that the fast decay of $p_X$ is due to the $ZZ$ interactions in this machine, 
and this can be suppressed by the proposed dynamical decoupling technique in Fig.~\ref{fig-DD}.

\section{Fault-tolerant logical zero-state encoding of the concatenated nine-qubit surface code}
\label{sec-concatenated}

The proposed encoding method is applicable only to the distance-three, nine-qubit surface code,
which is small for practical applications~\cite{Jones2012a,Fowler2012a,Gidney2021a}.
To extend the proposed method to larger codes, 
we consider the concatenation of the nine-qubit surface code with itself~\cite{Nielsen}, 
which leads to the distance-nine, 81-qubit code.

The logical zero state, $|0\rangle_{L2}$, encoded with the concatenated code can be generated straightforwardly
by the quantum circuit in Fig.~\ref{fig-encoder}\textbf{c}, 
where the physical zero states, physical Hadamard gates, and physical CNOT gates are, respectively, 
replaced by the encoded zero states $|0\rangle_{L1}$, 
encoded Hadamard gates, and encoded CNOT gates for the nine-qubit surface code, where 
the subscripts $L2$ and $L1$ denote the concatenation levels~\cite{Knill2005a,Goto2009a,Goto2013a}.
The fault-tolerant preparation of $|0\rangle_{L1}$ can be done by the proposed method explained in Sec.~\ref{sec-encoder}.
The encoded Hadamard gate can be implemented fault-tolerantly by the transversal Hadamard gate~\cite{Nielsen}, 
${H_1 H_2 H_3 H_4 H_5 H_6 H_7 H_8 H_9}$, followed by renumbering ($90^{\circ}$ rotating) the qubits as 
${1\to 7}$, ${2\to 6}$, ${3\to 1}$, ${4\to 2}$, ${9\to 3}$, ${8\to 4}$, ${7\to 9}$, and ${6\to 8}$.
The encoded CNOT gate can also be implemented fault-tolerantly by the transversal CNOT gate~\cite{Nielsen,Horsman2012a}.

To evaluate the performance of the straightforward encoding of $|0\rangle_{L2}$, 
we numerically simulate it and evaluate its logical error probability $p_{L2}$ using $Z$ measurement results.
This simulation is based on the stabilizer simulation~\cite{Aaronson2004a}.
In this simulation, we assume the error model where errors occur only in physical CNOT gates, 
which is modeled in a standard manner, that is, 
a physical CNOT gate is modeled by an ideal CNOT gate followed by 
one of 15 two-qubit Pauli errors with equal probability of $p_{\mathrm{CNOT}}/15$~\cite{Knill2005a,Goto2009a,Goto2013a}.
For the decoding of the concatenated code, 
we use soft-decision decoding based on conditional probability calculations~\cite{Goto2013a,Poulin2006a,Goto2014b} 
(see Appendix~\ref{appendix-soft-decoding}),
which is important to achieve almost optimal performance.

\begin{figure}
	\includegraphics[width=8cm]{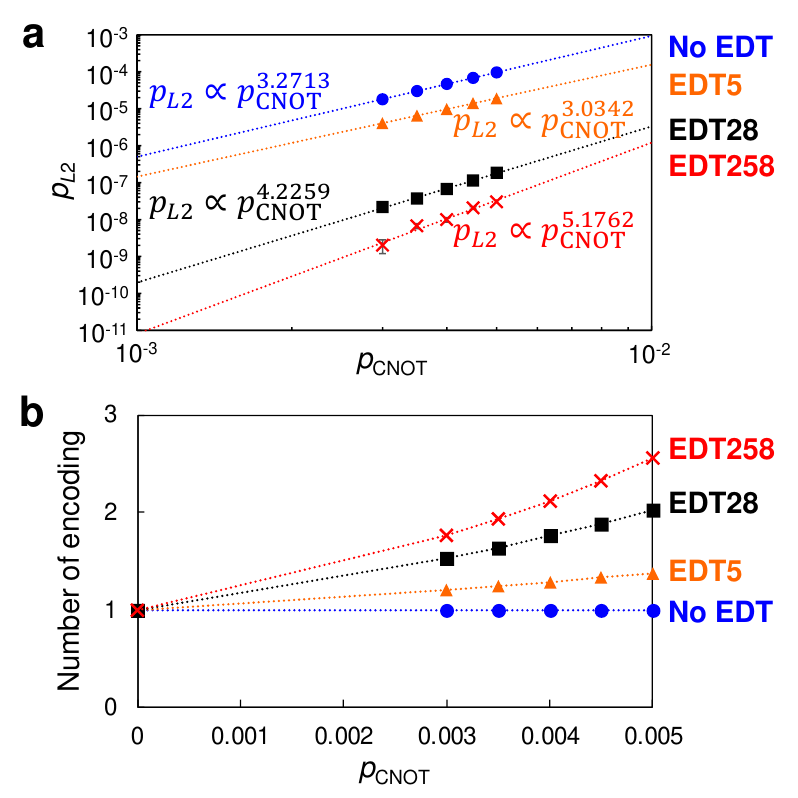}
	\caption{\textbf{Simulation results of the logical zero-state encoding of the concatenated code.}
	\textbf{a}~Logical error probabilities. 
	The circles, triangles, squares, and crosses show, respectively, the results for 
	encoding without error-detecting teleportations (EDTs), 
	with an EDT on the level-1 Qubit 5, 
	with EDTs on the level-1 Qubits 2 and 8, and
	with EDTs on the level-1 Qubits 2, 5, and 8.
	The dotted lines and proportionality relations show fitting results with a power function.
	The number of repetition to estimate the probabilities is $10^8$ for ``No EDT" and ``EDT5" 
	and $10^{10}$ for ``EDT28" and ``EDT258."
	\textbf{b}~Average total number of encoding until success. 
	The dotted lines are eye guides.
	In both \textbf{a} and \textbf{b}, statistical errors are negligible compared to symbol sizes, 
	except for ``EDT258" in \textbf{a}.}
	\label{fig-concatenate}
\end{figure}

The simulation results are shown by circles in Fig.~\ref{fig-concatenate}\textbf{a}.
Although the logical error probability can become lower than the physical error probability, 
its exponent of about 3 is smaller than that expected from the code distance of 9, namely, 5.
(In general, distance-nine codes can, in principle, correct four arbitrary independent qubit errors, 
leading to the exponent of 5.)

To improve the performance, error detection followed by postselection 
(restarting from the beginning if errors are detected until no errors are detected) 
is effective.
We perform level-1 error-detecting teleportations (EDTs)~\cite{Knill2005a,Goto2009a,Goto2013a} 
on the level-1 Qubits 2, 5, and/or 8 after the above straightforward generation of $|0\rangle_{L2}$.
The simulation results are shown in Fig~\ref{fig-concatenate}\textbf{a}.
When the EDT is performed only on Qubit 5, 
the logical error probability becomes lower, but the exponent is still about 3.
When the EDT is performed on Qubits 2 and 8, 
the logical error probability is reduced by two orders of magnitude, 
and also the exponent is increased to about 4 but still smaller than 5.
When the EDT is performed on Qubits 2, 5 and 8, 
not only the logical error probability becomes further lower, but also the exponent exceeds 5.
(Note that the hard-decision decoding for the concatenated code can, in principle, achieve the exponent up to 4, 
and therefore the present result achieving the exponent of 5 clearly shows the advantage of the soft-decision decoding.)
Thus, it has turned out that EDTs on Qubits 2, 5, and 8 are enough to achieve potential performance of the concatenated code.
In this case, the logical error probability is estimated around $10^{-11}$ for ${p_{\mathrm{CNOT}}=10^{-3}}$, 
which is sufficiently low for large-scale problems.
Figure~\ref{fig-concatenate}\textbf{b} also shows that 
the increase of overheads by the EDTs and postselection will not be very large.

\section{Conclusion}
\label{sec-discussion}

We have proposed a measurement-free fault-tolerant logical zero-state encoding method for  
the distance-three, nine-qubit surface code, and 
have experimentally demonstrated it using a superconducting quantum computer.
The experimental results indicate that the logical zero state was successfully prepared, and 
its logical error probability could be suppressed by error correction or detection.
We have also experimentally demonstrated that the dephasing due to residual $ZZ$ interactions in this machine 
can be suppressed by the proposed dynamical decoupling technique.
We have also proposed a fault-tolerant logical zero-state encoding method for the distance-nine, 81-qubit code 
obtained by concatenating the nine-qubit surface code with itself.
Because of the assumption of arbitrary two-qubit gates, 
this large-scale encoding is challenging for superconducting quantum computers, 
but will be possible for recently developed neutral-atom quantum computers with optical tweezers.
Thus, the present results will provide a new way to low-overhead fault-tolerant quantum computation.

\section*{Acknowledgments}
We acknowledge the use of IBM Quantum services for experiments in this work. 
The views expressed are those of the authors, and do not reflect the official policy or position of IBM or the IBM Quantum team.
The use of IBM's superconducting quantum computer on the cloud was supported by UTokyo Quantum Initiative.

\begin{appendix}

\section{Hard-decision decoding of the nine-qubit surface code}
\label{appendix-hard-decoding}

We estimate the $X$ errors in $|0\rangle_L$ as follows.
We first calculate the measurement results of the $Z$ stabilizers as
\begin{align}
s_{1} &= m_{6} m_{7}, 
\label{eq-sZ1}
\\
s_{2} &= m_{1} m_{2} m_{5} m_{6}, 
\\
s_{3} &= m_{4} m_{5} m_{8} m_{9}, 
\\
s_{4} &= m_{3} m_{4}, 
\label{eq-sZ4}
\end{align}
where $m_{j}$ denotes the measurement result of $Z_j$.
Next, we estimate the $X$ errors according to Table~\ref{table-hard-decoding}.
Finally, we flip the measurement results $\{ m_{j} \}$ according to the estimated $X$ errors.
Then, the value of $Z_L$ is estimated at $m_{1} m_{2} m_{3}$ after the flips.

\begin{table}[b]
	\caption{\textbf{$X$-error estimation from $Z$ measurement results.}
	Measurement results of the $Z$ stabilizers, $s_{1}$--$s_{4}$, are defined by Eqs.~(\ref{eq-sZ1})--(\ref{eq-sZ4}).}
	\begin{tabular}{ccccc} \hline \hline
	$s_{1}$ & $s_{2}$ & $s_{3}$ & $s_{4}$~ & $X$ errors
	\\ \hline 
	$+1$ & $+1$ & $+1$ & $+1$ & No error
	\\
	$-1$ & $+1$ & $+1$ & $+1$ & $X_7$
	\\
	$+1$ & $-1$ & $+1$ & $+1$ & $X_2$
	\\
	$+1$ & $+1$ & $-1$ & $+1$ & $X_8$
	\\
	$+1$ & $+1$ & $+1$ & $-1$ & $X_3$
	\\
	$-1$ & $-1$ & $+1$ & $+1$ & $X_6$
	\\
	$-1$ & $+1$ & $-1$ & $+1$ & $X_7 X_8$
	\\
	$-1$ & $+1$ & $+1$ & $-1$ & $X_3 X_7$
	\\
	$+1$ & $-1$ & $-1$ & $+1$ & $X_5$
	\\
	$+1$ & $-1$ & $+1$ & $-1$ & $X_2 X_3$
	\\
	$+1$ & $+1$ & $-1$ & $-1$ & $X_4$
	\\
	$-1$ & $-1$ & $-1$ & $+1$ & $X_5 X_7$
	\\
	$-1$ & $-1$ & $+1$ & $-1$ & $X_3 X_6$
	\\
	$-1$ & $+1$ & $-1$ & $-1$ & $X_4 X_7$
	\\
	$+1$ & $-1$ & $-1$ & $-1$ & $X_3 X_5$
	\\
	$-1$ & $-1$ & $-1$ & $-1$ & $X_4 X_6$
	\\ \hline \hline
	\end{tabular}
	\label{table-hard-decoding}
\end{table}

\begin{widetext}

\section{Soft-decision decoding of the concatenated nine-qubit surface code}
\label{appendix-soft-decoding}

First, the soft-decision decoding rule for the nine-qubit surface code is given as follows~\cite{Goto2013a,Goto2014b}:
\begin{align}
P^{(1)}(z) &= \frac{R^{(1)}(z)}{R^{(1)}(1)+R^{(1)}(-1)},
\\
R^{(1)}(z)
&=
\sum_{z_1=\pm 1} 
\sum_{z_2=\pm 1} 
\sum_{z_3=\pm 1} 
\sum_{z_4=\pm 1} 
\sum_{z_5=\pm 1} 
\sum_{z_6=\pm 1} 
\sum_{z_7=\pm 1} 
\sum_{z_8=\pm 1} 
\sum_{z_9=\pm 1}
P_1^{(0)}(z_1) 
P_2^{(0)}(z_2) 
P_3^{(0)}(z_3) 
P_4^{(0)}(z_4) 
P_5^{(0)}(z_5) 
P_6^{(0)}(z_6) 
\nonumber
\\
&\times 
P_7^{(0)}(z_7) 
P_8^{(0)}(z_8) 
P_9^{(0)}(z_9) 
\delta ({z_1 z_2 z_3 = z})
\delta ({z_6 z_7 = 1})
\delta ({z_1 z_2 z_5 z_6 = 1})
\delta ({z_4 z_5 z_8 z_9 = 1})
\delta ({z_3 z_4 = 1}),
\end{align}
where $P^{(1)}(z)$ is the probability that the level-1 $Z_L$ has the value of $z$, $R^{(1)}(z)$ is the corresponding relative probability, 
$z_j$ denotes the correct value of $Z_j$ for the $j$th physical qubit, $P_j^{(0)}(z_j)$ is the corresponding probability, 
and the $\delta (\textrm{condition})$ is the indicator function 
taking 1 if the condition is true and otherwise 0. 
Assuming equal physical-qubit error probability of $p_e$, 
we have
\begin{align}
P_j^{(0)}(m_j)=1-p_e,~P_j^{(0)}(-m_j)=p_e,
\end{align}
where $m_j$ is the measurement result of $Z_j$.
In this work, we set $p_e$ to 0.01. 
(The soft-decision decoding is insensitive to the setting of $p_e$~\cite{Goto2013a}.)
Thus, we can obtain $P^{(1)}(z)$ using the measurement results $\{ m_j \}$.

Next, the soft-decision decoding rule for the concatenated code is given similarly as follows~\cite{Goto2013a,Goto2014b}:
\begin{align}
P^{(2)}(z) &= \frac{R^{(2)}(z)}{R^{(2)}(1)+R^{(2)}(-1)},
\\
R^{(2)}(z)
&=
\sum_{z_1=\pm 1} 
\sum_{z_2=\pm 1} 
\sum_{z_3=\pm 1} 
\sum_{z_4=\pm 1} 
\sum_{z_5=\pm 1} 
\sum_{z_6=\pm 1} 
\sum_{z_7=\pm 1} 
\sum_{z_8=\pm 1} 
\sum_{z_9=\pm 1}
P_1^{(1)}(z_1) 
P_2^{(1)}(z_2) 
P_3^{(1)}(z_3) 
P_4^{(1)}(z_4) 
P_5^{(1)}(z_5) 
P_6^{(1)}(z_6) 
\nonumber
\\
&\times 
P_7^{(1)}(z_7) 
P_8^{(1)}(z_8) 
P_9^{(1)}(z_9) 
\delta ({z_1 z_2 z_3 = z})
\delta ({z_6 z_7 = 1})
\delta ({z_1 z_2 z_5 z_6 = 1})
\delta ({z_4 z_5 z_8 z_9 = 1})
\delta ({z_3 z_4 = 1}),
\end{align}
\end{widetext}
where $P^{(2)}(z)$ is the probability that the level-2 $Z_L$ has the value of $z$, $R^{(2)}(z)$ is the corresponding relative probability, 
$z_j$ denotes the correct value of the level-1 $Z_L$ for the $j$th level-1 encoded qubit, $P_j^{(1)}(z_j)$ is the corresponding probability 
obtained from the measurement results as explained above.

Thus, we estimate the value of the level-2 $Z_L$ at 1 if  ${P^{(2)}(1) > P^{(2)}(-1)}$, otherwise at $-1$.

\section{Fidelity estimation of the experimentally prepared $|0\rangle_L$}
\label{appendix-fidelity}

The fidelity of the experimentally prepared $|0\rangle_L$ is formuated as 
${F_0 = {_L \langle 0|\rho_0 |0 \rangle_L}}$, 
where $\rho_0$ denotes the density operator describing the experimentally prepared $|0\rangle_L$.
In the following, we show that ${F_0 \ge p_Z + p_X - 1}$.

Consider two bits, $b_Z$ and $b_X$, defined as 
${b_Z = 0}$ if all the five $Z$ stabilizers of $|0\rangle_L$ are 1, otherwise ${b_Z = 1}$, and 
${b_X = 0}$ if all the four $X$ stabilizers of $|0\rangle_L$ are 1, otherwise ${b_X = 1}$. 
We also denote the probability of $(b_Z,b_X)$ by $P_b(b_Z,b_X)$.
Then, we have ${F_0=P_b(0,0)}$, ${p_Z=P_b(0,0)+P_b(0,1)}$, and ${p_X=P_b(0,0)+P_b(1,0)}$.
We also have ${P_b(0,0)+P_b(1,0)+P_b(0,1)+P_b(1,1)=1}$.
Using these relations, we can easily obtain the desired result:
\begin{align}
F_0 = p_Z + p_X - 1 + P_b(1,1) \ge p_Z + p_X - 1.
\end{align}

\end{appendix}


\begin{thebibliography}{99}
\bibitem{comment-stabilizer}
An $n$-qubit stabilizer state is described as a simultaneous eigenstate of $n$ stabilizers with eigenvalues of 1~\cite{Nielsen}.
\bibitem{comment-correction}
The $X_2 X_3$ error can be found by the flips of the two $Z$ stabilizers,  $Z_1 Z_2 Z_5 Z_6$ and $Z_3 Z_4$, and 
the $X_7 X_8$ error can also be found by the flips of $Z_6 Z_7$ and $Z_4 Z_5 Z_8 Z_9$. 
Also, there are no correlated $Z$ errors, because $|0\rangle_L$ has another stabilizer $Z_L$ 
and consequently the weights of its $Z$ stabilizers are still two or three after Step~2, 
as shown in Fig.~\ref{fig-encoder}\textbf{d}.
\bibitem{comment-T2}
The larger value of $T_2$ than $2T_1$ for Qubit~23 is due to a calibration error in this system.
\end{thebibliography}
\end{document}